\documentclass{article}
\usepackage{spconf,amsmath,graphicx}
\usepackage[nolist,nohyperlinks]{acronym}
\usepackage{algorithm}
\usepackage{algorithmic}
\usepackage{amssymb}
\usepackage{hyperref}

\acrodef{MSE}[MSE]{\emph{Mean-Square Error}}
\acrodef{ATC}[ATC]{Adapt-Then-Combine}
\acrodef{CTA}[CTA]{Combine-Then-Adapt}
\acrodef{AVID}[AVID]{Adaptive Velocity and Intermediate Distance}
\acrodef{TC}[TC]{Tangent-Chord} % Include the acronyms.tex file

\title{Crowd Modeling and Control via Cooperative Adaptive Filtering}
\name{Zirui Wan and Saeid Sanei\thanks{This work has been submitted to the IEEE for possible publication. Copyright may be transferred without notice, after which this version may no longer be accessible.}}
\address{Department of Electrical and Electronic Engineering, Imperial College London, UK}

\begin{document}
%\ninept
%

\maketitle

\begin{abstract}

This paper introduces a crowd modeling and motion control approach that employs diffusion adaptation within an adaptive network. In the network, nodes collaboratively address specific estimation problems while simultaneously moving as agents governed by certain motion control mechanisms. Our research delves into the behaviors of agents when they encounter spatial constraints. Within this framework, agents pursue several objectives, such as target tracking, coherent motion, and obstacle evasion. Throughout their navigation, they demonstrate a nature of self-organization and self-adjustment that drives them to maintain certain social distances with each other, and adaptively adjust their behaviors in response to the environmental changes. Our findings suggest a promising approach to mitigate the spread of viral pandemics and averting stampedes.

\end{abstract}
\section{Introduction}
\label{sec:intro}

\subsection{Background}
Crowd modeling and control has become an essential research field with widespread applications in urban planning, public safety, and public health \cite{yangreview,gayathri2017review,echeverria2021estimating}. As the world grapples with challenges such as pandemics and large-scale public gatherings, the importance of understanding and effectively managing crowd dynamics has never been more pronounced.

One critical aspect of crowd modeling is ensuring that individual agents, representing people within the crowd, maintain a safe distance from each other \cite{heylighen2001science}. An effective solution to this challenge is inspired by the self-organization behavior found in various natural phenomena such as fish schools \cite{fish}, bird flocks \cite{bird}, and honey bee swarms \cite{bee}. Instead of being governed by a predetermined, global blueprint, their coherent motion emerges from repeated local interactions. Some prior studies have successfully simulated self-organization behaviors by utilizing mobile adaptive networks \cite{fishschool1,fishschool2,fishschool3,bacterial2}.  In these networks, nodes function as agents, and diffusion adaptation serves as a distributed learning strategy.

In this paper, we expand beyond self-organization to introduce self-adjustment into our crowd model. Specifically, the agents can make real-time responses according to the spatial constraints in their immediate surroundings. This enables the formation of more realistic crowd movement in predefined environments, such as a corridors with different widths. We have examined the model through simulations and the results reveal that our approach effectively emulates crowd behaviors. Such a modeling approach can significantly contribute to reducing the risk of pandemic diseases and stampedes.

\subsection{Diffusion Adaptation}
\label{sec:format}

We first consider the problem of distributed estimation. In our networks, nodes collaborate to estimate certain vectors of interest by solving global optimization problems. Specifically, the objective is to determine the location of a target $w$ and the velocity of the agents' centre of gravity $v^g$, which facilitate target pursuit and coordinated movement across the nodes, respectively \cite{fishschool2}. To accomplish this, diffusion adaptation is utilized as the distributed learning strategy. This approach enables nodes to learn through local interactions and diffuse the local information, enhancing overall performance.

To solve the distributed optimization issue, a general measurement model and the cost function is given as \cite{diffusion}:
\begin{equation}
d_k(i)=u_{k, i} w^{o}+n_k(i)
\label{equ:measurement relation}
\end{equation}
\begin{equation}
J^{glob}(w)=\sum_{k=1}^N \mathbb{E}\left|d_k(i)-u_{k, i} w\right|^2
\label{equ:Jglob_mse}
\end{equation}
where $w^{o}$ is the desired vector, and at each time instant $i$, every node $k$ acquires a measurement $d_k(i)$, along with a row regression vector $u_{k,i}$. $n_k(i)$ symbolizes a zero-mean white random noise. The global cost function \eqref{equ:Jglob_mse} minimizes the summation of individual Mean-Square Error cost functions over $N$ nodes.

Two solutions are given in \cite{diffusion}, which are \ac{ATC} and \ac{CTA}, in this work, we employ the \ac{ATC} strategy as follows:
\begin{equation}
\begin{aligned}
\psi_{k, i} & =w_{k, i-1}+\mu_k  u_{k, i}^*\left[d_{k}(i)-u_{k, i} w_{k, i-1}\right] \\
w_{k, i} & =\sum_{\ell \in \mathcal{N}_k} a_{\ell k} \psi_{\ell, i}
\end{aligned}
\label{equ:ATC}
\end{equation}
where $\mathcal{N}_k$ is the neighborhood of node $k$, which consists of nodes within certain radius $R$. $\mu_k$ is the adaptation step size, and $a_{\ell k}$ is the combination factor which satisfies: 
\begin{equation}
\label{alk}
\sum_{\ell=1}^N a_{\ell k}=1, \, a_{\ell k} \geq 0, \, \text { and } \, a_{\ell k}=0 \, \text { if } \, \ell \notin \mathcal{N}_k
\end{equation}

\section{Methodology}
\subsection{Estimating \texorpdfstring{$w$}{Lg} and \texorpdfstring{$v^g$}{Lg}}
To address our specific estimation problem, we first define a set of measurements: $d_k(i)$, $p_{k,i}$, $x_{k,i}$ and $v_{k,i}$. These variables represent, at time instant $i$, the distance between node $k$ and the position of a target for node $k$, the unit direction vector pointing from node $k$ to the target, the location of node $k$ and the velocity of node $k$, respectively. The diffusion adaptation strategy for the estimation of $w$ and $v^g$ is shown in Algorithm \ref{alg:diffuse adapt} \cite{fishschool2}. Here, $\mu_k$ and $\nu_k$ are step sizes, while $a_{\ell k}^w$ and $a_{\ell k}^v$ are combination factors for the combination step.
\begin{algorithm}
\caption{Diffusion Adaptation Algorithm \cite{fishschool2}}\label{alg:diffuse adapt}
\begin{algorithmic}[0] % The argument [1] adds line numbers
% \REQUIRE Each node has measurements $\{ d_k(i),p_{k,i},x_{k,i},v_{k,i} \}$.
\FOR{$k = 1$ to $N$}
    \STATE Each node performs information exchange and adaptations simultaneously.
    \begin{equation}
    \begin{aligned}
    \begin{cases}
    \psi_{k, i} & =w_{k, i-1}+\mu_k  p_{k, i}^T \left[d_k(i)+p_{k, i}(x_{k, i}-w_{k, i-1})\right] \\
    \phi_{k, i} & =v_{k, i-1}^g+\nu_k \left(v_{k, i}-v_{k, i-1}^g\right)\\   
    \end{cases}
    \end{aligned}
    \end{equation}
    \STATE Each node combines intermediate estimates from its neighbourhood:
    \begin{equation}
    \begin{aligned}
    \begin{cases}
    w_{k, i} & =\sum_{\ell \in \mathcal{N}_k} a_{\ell k}^w \psi_{\ell, i}\\
    v_{k, i}^g & =\sum_{\ell \in \mathcal{N}_k} a_{\ell k}^v \phi_{\ell, i}\\
    \end{cases}
    \end{aligned}
    \end{equation}
\ENDFOR
% \RETURN Each node ends up with $w_{k, i}$ and $v_{k, i}^g$. 
\end{algorithmic}
\end{algorithm}

\subsection{Motion Control Mechanism}
\label{sec:pagestyle}
Our nodes are moving in a two-dimensional space, $\mathbb{R}^2$. The position of each node is updated over discrete time instants, with the updating equation as follows:
\begin{equation}
\label{equ:Mupdate}
x_{k, i+1}=x_{k, i}+\Delta t \cdot v_{k, i+1}
\end{equation}
where $\Delta t$ is the time step for each update. The various desires of the agents, such as pursuing the target, maintaining coherent motion, and avoiding obstacles, are all reflected in their velocity components and resulting different movement trajectories. 

\begin{figure}
\centering
\includegraphics[width=7cm,height=6.5cm]{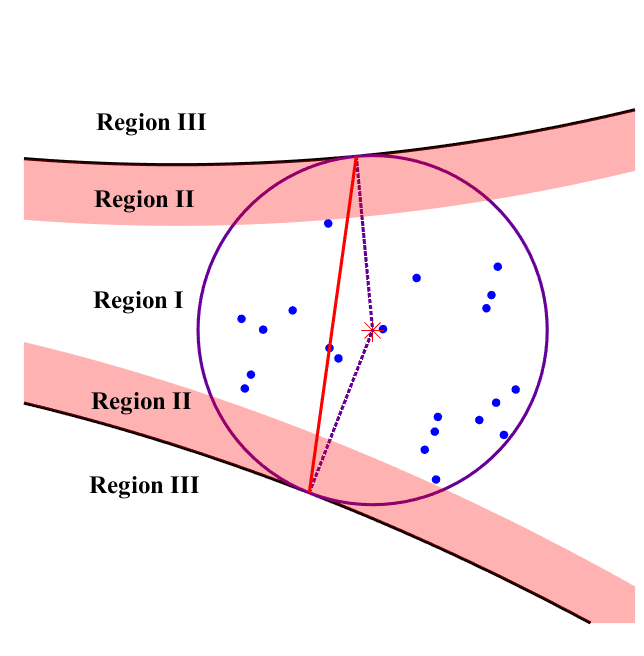}
\caption{A diagram of nodes in a corridor. The blue dots represent the nodes, while the red asterisk denotes the center of the tangent circle. The Tangent-Chord of a node is marked by a red line.}\label{fig:diagram}
\end{figure}

Two factors influence the velocity of the nodes within the network. The first factor is the location of the nodes. In Fig. \ref{fig:diagram}, we can see that the space is divided into three regions. If a node is located in Region I, its primary goal is to pursue the target since the spatial constraints are still far away from it. However, if the node comes too close to spatial constraints, situated in Region II, such that its distance from the wall or other obstacles is less than a certain tolerable distance $d$, the primary objective shifts to navigating around the obstacles to avoid collision. Additionally, if the node is on a trajectory that would lead it out of the wall boundaries to region III, it will remain stationary at the current time instant, ensuring compliance with spatial limitations. The velocity component $v^a$ is given by:
\begin{equation}
    \begin{aligned}
        &v_{k, i+1}^a=\frac{w_{k, i}-x_{k, i}}{\left\| w_{k, i}-x_{k, i} \right\|} \quad(\text { Region I }) \\
        &v_{k, i+1}^a=\frac{1}{2} \Bigl[ \frac{w_{k, i}-x_{k, i}}{\left\| w_{k, i}-x_{k, i} \right\|} \\
        &+\eta \left( d-\left\|x_{k,i}-s_{k,i} \right\| \right) \left( \frac{x_{k,i}-s_{k,i}}{\left\|x_{k,i}-s_{k,i} \right\|}\right) \Bigr] \quad (\text { Region II })\\
        &v_{k, i+1}^a=0 \quad (\text { Region III })
    \end{aligned}
\end{equation}
where  $s_{k,i}$ represents the position of the nearest point on an obstacle relative to node $k$, and $\eta$ is a scalar for the obstacle avoidance component, since sometimes nodes tend to escape from the obstacles with a higher velocity to avoid collision. 

Another factor is the desire of coherent motion, where nodes aim to maintain a uniform speed and sustain a certain intermediate distance from each other. To achieve this, we can employ a mechanism modeled for the behavior of fish schools \cite{fishschool2,fishschool3}. The equation for the estimation of this velocity component is as follows:
\begin{equation}
    v_{k,i+1}^b=(1-\lambda)v_{k,i}^g +\gamma \delta_{k,i}
\end{equation}
\begin{equation}
    \delta_{k,i}=\frac{1}{\left|\mathcal{N}_k\right|-1} \sum_{\ell \in \mathcal{N}_k \backslash\{k\}}  \left(1-\frac{r_{k,i}}{\left\|x_{\ell,i}-x_{k,i}\right\|}\right) \left(x_{\ell,i}-x_{k,i}\right)
\end{equation}
where $r_{k,i}$ is an adaptive desired intermediate distance the nodes want to maintain, while $(1-\lambda)$ and $\gamma$ are non-negative weight factors. The velocity components can be systematically organized into an updating equation, as detailed in:
\begin{equation}
\label{equ:v_upd}
\begin{aligned}
    v_{k,i+1}&=\lambda\alpha_{k,i}v_{k,i+1}^{a}+ v_{k,i+1}^{b}\\
            &=\lambda\alpha_{k,i}v_{k,i+1}^{a}+(1-\lambda)v_{k,i}^{g}+\gamma \delta_{k,i}
\end{aligned}
\end{equation}
where $\alpha_{k,i}$ denotes an adaptive weight factor which controls the moving speed of nodes.

\subsection{Adaptive Velocity and Intermediate Distance}
\label{sec:typestyle}
To make our model more realistic, we must take into account the behavior of crowd when encountering narrow passages. In real-life scenarios, groups of agents often navigate corridors that exhibit variations in width, with the narrowest segment referred to as the ``neck" of the corridor. To traverse through the corridor's neck efficiently and minimize collisions between the nodes and corridor walls, the network must adjust their configurations. Specifically, nodes should move relatively quickly through narrow area and slow down in wider area. This can be achieved by adjusting the velocity parameter $\alpha_{k,i}$. Additionally, to further prevent collisions with the walls, nodes are allowed to reduce their desired distance $r_{k,i}$, to a reasonable value at the neck. We refer to this behavior as \ac{AVID}.

A critical problem that must be addressed to implement \ac{AVID} is measuring the corridor width for a node at a specific position. Here, we employ a novel technique named \ac{TC}. To be precise, for each agent $k$, we first construct a circle that is situated inside the corridor, tangent to both walls, and the agent falls on the chord between the two tangent points. Then, this chord's length serves as a measure of the corridor's width at that specific location of the node. An illustration of \ac{TC} is given in Fig. \ref{fig:diagram}. Since in our model, the functions defining the walls are assumed known, we can obtain the width estimates $l_{k,i}^c$ by solving a system of equations.

To establish the mechanism of \ac{AVID}, we firstly introduce some new parameters and variables. We define $l^{s}$ as the standard width of the corridor. This value serves as a threshold: only when $l_{k,i}^c$ is less than $l^{s}$, will the nodes consider modifying their velocities and intermediate distances. Since our goal is to increase the velocity while decreasing the intermediate distance, we establish a minimum distance $r_{min}$ and a maximum velocity parameter $\alpha_{max}$, as well as standard reference terms $r$ and $\alpha$. The mechanism is given in Algorithm \ref{alg:adaptive velocity and distance}.
\begin{algorithm}
\caption{\ac{AVID} Algorithm}\label{alg:adaptive velocity and distance}
\begin{algorithmic}[0] % The argument [1] adds line numbers
% \REQUIRE Each node has a measurement $l_{k,i}^c$, and $\{l^s,r_{min},\alpha_{max},r,\alpha\}$ are predefined.
\FOR{$k = 1$ to $N$}
    \IF{$l_{k,i}^c$ $<$ $l^{s}$}
    \STATE Each node updates $\alpha_{k,i}$ and $r_{k,i}$ according to $l_{k,i}^c$.

        \begin{equation}
        \begin{aligned}
            r_{k,i}&=\left(1-\frac{l_{k,i}^c}{l^s} \right) r_{min}+\left(\frac{l_{k,i}^c}{l^s}\right) r\\
            \alpha_{k,i}&=\left(\frac{l_{k,i}^c}{l^s} \right) \alpha+\left(1-\frac{l_{k,i}^c}{l^s}\right) \alpha_{max}
        \end{aligned}
       \end{equation}
        
    \ELSE
    \STATE Each node takes the standard terms.
        \begin{equation}  
        \begin{aligned}
            \alpha_{k,i}&=\alpha\\
            r_{k,i}&=r
        \end{aligned}
        \end{equation}
    \ENDIF
\ENDFOR
% \RETURN Each node ends up with $\alpha_{k,i}$ and $r_{k,i}$.
\end{algorithmic}
\end{algorithm}

\section{Simulation Results and Discussion}
\label{sec:majhead}

In this section, we demonstrate the simulation results of agents in a predefined corridor. The parameters involved are specified as follows. The time step $\Delta t$ is set to $0.5$, while neighborhood radius $R=3.5$ and standard desired distance $r=3$. The tolerable distance $d$ is set to $2$. For \ac{AVID}, the coefficients are $r_{min}=2$, $\alpha_{max}=4$, $l_s=16$. All step sizes are set to $0.5$. The weight factors are $\lambda=0.5$ and $\alpha=\gamma=\eta=2$. The corridor's upper and lower walls are defined by two simple quadratic curves (in general they can be any continuous functions):
\begin{equation}
    \begin{cases}
        \begin{aligned}
            f^{upper}(x)&=0.008(x-10)^2+20\\
            f^{lower}(x)&=0.008(x+10)^2+14
        \end{aligned}
    \end{cases}
    \label{equ:walls}
\end{equation}
To analyze the performance of nodes' mobility in the corridor model, we define four variables: $v^{mean}$, $r^{mean}$, $N^{obs}$ and $N^{neck}$.
\begin{equation}
    v_i^{mean}\stackrel{\triangle}{=} \frac{1}{N} \sum_{k=1}^N\left\|v_{k,i}\right\|
\end{equation}
\begin{equation}
    r_i^{mean}\stackrel{\triangle}{=} \frac{1}{N} \sum_{k=1}^N \frac{1}{\left|\mathcal{N}_k\right|-1} \sum_{\ell \in \mathcal{N}_k \backslash\{k\}}\left\|x_{\ell,i}-x_{k,i}\right\|
\end{equation}
where $v^{mean}$ represents the average velocity magnitude across all nodes in the network, providing a measure of the overall speed of the system. Meanwhile, $r^{mean}$ is calculated as the average of all the distances between neighboring nodes. $N^{obs}$  denotes the number of nodes situated in region II. Those obstructed nodes must adjust their directions to avoid collisions. Finally, $N^{neck}$ signifies the number of nodes located at the corridor's neck, serving as a measure of the rate of passage through this constrained area. 

\begin{figure}[htb]
\centering
\includegraphics[width=8cm]{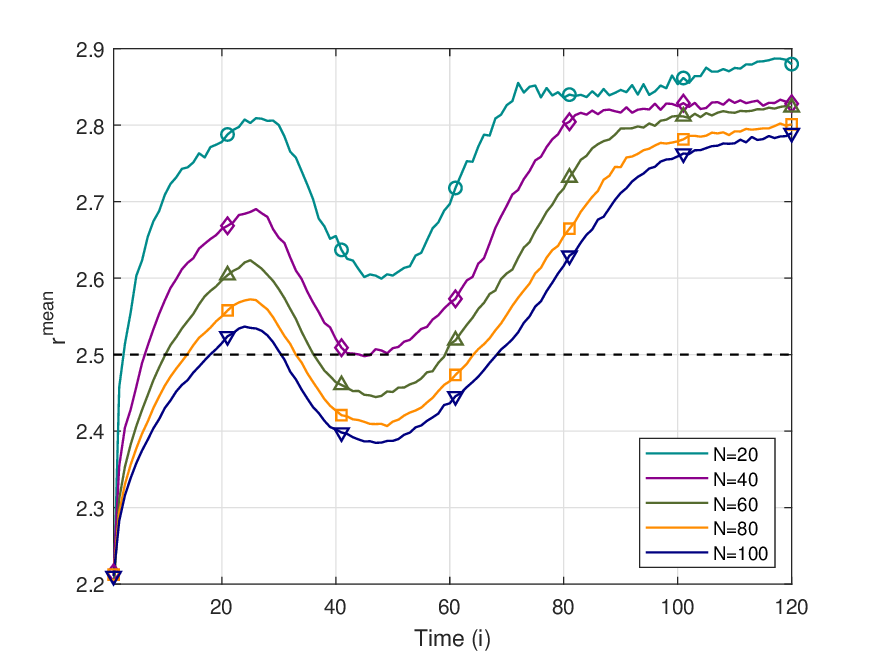}
\caption{$r^{mean}$ for nodes of different quantities over $120$ iterations.}\label{fig:Npassage}
\end{figure}

Fig. \ref{fig:Npassage} illustrates the $r^{mean}$ for different quantities of agents passing through the corridor. While the distances between agents are reduced to certain degrees at the neck, those with higher quantities of nodes exhibit more pronounced decreases in $r^{mean}$. This observation has practical implications: for example, to minimize risks such as virus transmission and stampede accidents, a safe distance of $2.5$ units must be maintained. Consequently, any population consisting of more than $40$ agents should be either diverted or issued a warning.

\begin{figure}
\centering
\includegraphics[width=9cm,height=7cm]{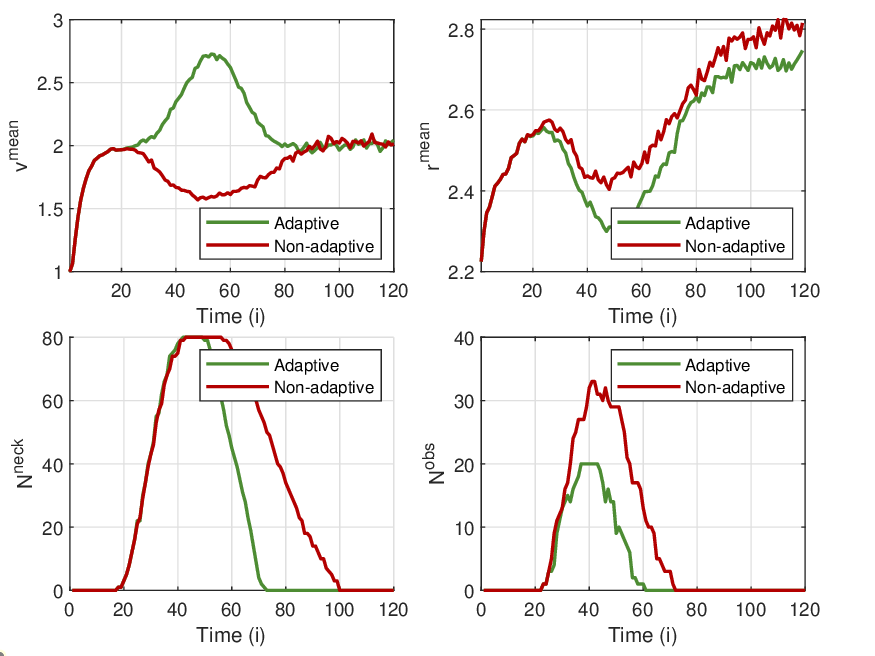}
\caption{A comparison between adaptive and non-adaptive nodes on $v^{mean}$, $r^{mean}$, $N^{obs}$ and $N^{neck}$.}\label{fig:urNN}
\end{figure}

Fig. \ref{fig:urNN} compares the results of $v^{mean}$, $r^{mean}$, $N^{obs}$ and $N^{neck}$ between adaptive (\ac{AVID} implemented) and non-adaptive configurations. The image for $v^{mean}$ reveals that without adaptive velocity, a subtle decline in $v^{mean}$ is noticeable when going through the corridor's neck, as some nodes slow down or come to a stop to avoid collisions. When \ac{AVID} is implemented, there is a pronounced increase in speed as the nodes approach the neck. Similar phenomenon can be observed with regard to the mean distance $r^{mean}$. The adaptive case reveals a further reduced $r^{mean}$ at the neck, facilitating smoother movement through the constrained space and minimizing the potential for collisions. The results for $N^{obs}$ and $N^{neck}$ corroborate these findings, confirming fewer collisions and more rapid passage through the corridor. Therefore, \ac{AVID} gives a more realistic crowd modeling which makes a compromise between the speed and distance maintenance ability.

\begin{figure}[h]
\centering
\includegraphics[width=9cm,height=8cm]{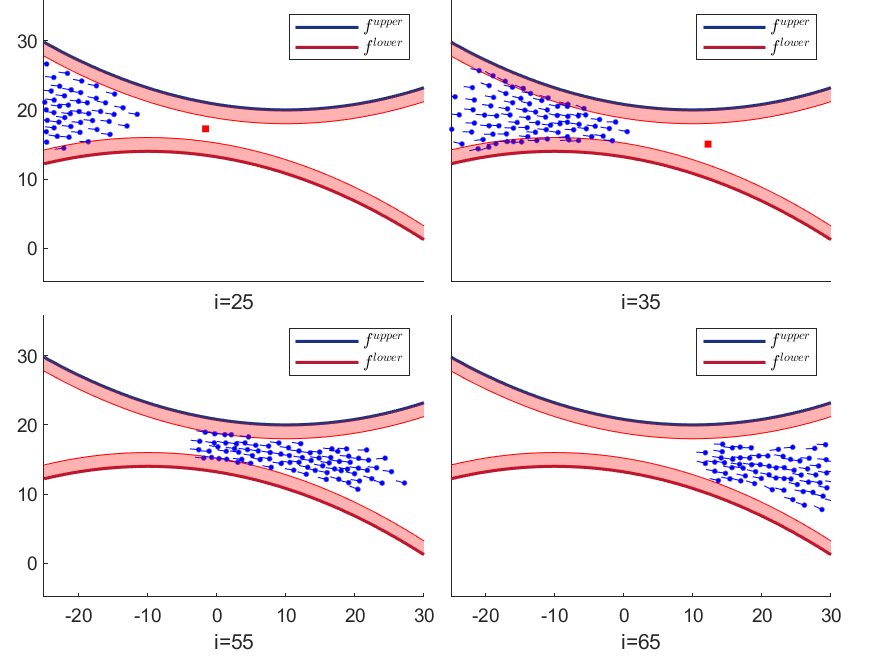}
\caption{The maneuver diagram captured from four distinct time instants.}\label{fig:passage}
\end{figure}

The maneuver of agents in the corridor is depicted in Fig. \ref{fig:passage}. The symbol ``$\bullet$" denotes the agents and their tails ``-" represents their moving directions. The red symbol ``$\blacksquare$" is a moving target. Both self-adjustment and self-organization are showcased through the formation contraction and restoration, respectively. A demonstration video including simulation results under various scenarios can be found at the link: \href{https://youtu.be/oxqP0riGX8E}{https://youtu.be/oxqP0riGX8E}.
% At time instant $i=35$, the leading nodes are seen to contract, due to both the force of obstacle avoidance and reduced $r_{k,i}$. As the formation reaches $i=50$, it is squeezed into a narrow and elongated shape, indicating the self-adjustment nature of our model. Upon leaving the neck, at $i=65$, the nodes gradually revert to their original formation, demonstrating their self-organization ability. 

\section{Conclusions}
\label{sec:majhead}
In this paper, we present a crowd modeling and control method using cooperative adaptive filtering. In this model, agents exhibit significant self-adjustment and self-organization abilities, ensuring that agents navigate spatial constraints both safely, by maintaining an appropriate distance, and efficiently, by adjusting velocity and intermediate distances. This work holds potential to mitigate the risk of pandemic infections and prevent stampedes. The information about the agent locations can be easily captured from our mobile phones or AirTags in real-life applications and used for example to automatically navigate vehicles and wheelchairs in a complex environment.
\vfill
\pagebreak

\bibliographystyle{IEEEbib}
\bibliography{main}

\end{document}